\documentclass[ reprint, amsmath,amssymb,aps,
]{revtex4-1}
\usepackage{graphicx}
\usepackage{color}
\usepackage{amsmath}
\usepackage{amssymb}
\usepackage{amsthm}
\usepackage{booktabs}
\usepackage[dvipdfm,colorlinks,urlcolor=blue,linkcolor=blue,anchorcolor=blue,citecolor=blue]{hyperref}

\begin{document}

\title{Universal law of thermalization for one-dimensional perturbed Toda lattices}

\author{Weicheng Fu}
\author{Yong Zhang}
\email{yzhang75@xmu.edu.cn}
\author{Hong Zhao}

\affiliation{Department of Physics and Jiujiang Research Institute, Xiamen University, Xiamen 361005, Fujian, China}

\date{\today }
\begin{abstract}
  The Toda lattice is a nonlinear but integrable system.~Here we study the thermalization problem in one-dimensional, perturbed Toda lattices in the thermodynamic limit.~We show that the thermalization time, $T_{eq}$, follows a universal law; i.e.,~$T_{eq}\sim \epsilon^{-2}$, where the \textit{perturbation strength}, $\epsilon$, characterizes the nonlinear perturbations added to the Toda potential.~This universal law applies generally to weak nonlinear lattices due to their equivalence to perturbed Toda systems.
\end{abstract}

\maketitle

\section{\label{sec:1}Introduction}

The problem of thermalization in weak nonlinear systems has a long studying history but is still far from being resolved.~The first numerical experiment, aiming to observe the rates of mixing and thermalization in a system of reversible microscopic dynamics, was accomplished in the 1950's by Fermi, Pasta, Ulam (FPU)~\cite{Fermi1955}, and Tsingou~\cite{Fermi1955, dauxois:ensl-00202296}. Their numerical results showed very little tendency toward equipartition of energy among the degrees of freedom, which is known as the FPU recurrence~\cite{Fermi1955, dauxois:ensl-00202296, PhysRevLett.98.047202, PhysRevX.4.011054, PhysRevLett.117.163901, PhysRevX.7.011025, NaturePhotonics.12.303, PhysRevX.8.041017, PhysRevLett.15.240}. This seminal work failed to observe the expected picture but opened up two entirely new research fields: nonlinear science~\cite{PhysRevLett.15.240, Dauxois:2006zz, PhysRevLett.19.1095, 1966SPhD1130I, CHIRIKOV1979263, doi:10.1063/1.1889345, doi:10.1063/1.1855036, doi:10.1063/1.1849131} and computational science~\cite{doi:10.1063/1.1861554,Porter2009Fermi}.

Many efforts have been made to explain FPU's results~\cite{Fermi1955, dauxois:ensl-00202296, PhysRevLett.98.047202, PhysRevX.4.011054, PhysRevLett.117.163901, PhysRevX.7.011025, NaturePhotonics.12.303, PhysRevX.8.041017, PhysRevLett.15.240, Dauxois:2006zz, PhysRevLett.19.1095, 1966SPhD1130I, CHIRIKOV1979263, doi:10.1063/1.1889345, doi:10.1063/1.1855036, doi:10.1063/1.1849131, doi:10.1063/1.1861554, Porter2009Fermi, PhysRevE.58.7165, PhysRevLett.95.064102, PhysRevE.73.036618, rangarajan1998kolmogorov-arnold-moser, 2008LNP728G, PhysRevA.1.59, TUCK1972399, BIVINS197365, SHOLL1990253, FORD1992271, doi:10.1063/1.1858115, doi:10.1063/1.1861264, PhysRevE.95.060202} and many theories have been put forward for understanding the relaxation problem, e.g., the soliton theory~\cite{PhysRevLett.15.240}, the Chirikov resonance overlap theory~\cite{1966SPhD1130I, CHIRIKOV1979263}, the mode-coupling theory~\cite{PhysRevE.58.7165}, the $q$-breathers theory~\cite{PhysRevLett.95.064102, PhysRevE.73.036618}, as well as the Kolmogorov-Arnold-Moser theorem~\cite{rangarajan1998kolmogorov-arnold-moser, 2008LNP728G} enunciated by Kolmogorov at nearly the same time of FPU's work.~However, the original goal of the FPU problem, that is, to answer whether a simple dynamical system can reach the thermalized state at arbitrarily weak nonlinearity and what properties the equipartition process may have, have not been achieved.~To this end, some studies are inconsistent with each other~\cite{2008LNP728G}.~For instance, for the FPU-$\beta$ model, while Berchialla {\it et al.}~showed that the equipartition time $T_{eq}$ depends on the energy density $\varepsilon$ in a stretched exponential law in the thermodynamic limit, i.e., $T_{eq}\sim \exp(-\varepsilon^{1/4})$~\cite{BERCHIALLA2004167}, DeLuca {\it et al.}~suggested a power-law relationship $T_{eq}\sim \varepsilon^{-3}$ instead~\cite{PhysRevE.51.2877, *PhysRevE.60.3781}. Benettin {\it et al.} indicated however a crossover from the stretched exponential law, $T_{eq}\sim \exp(-\varepsilon^{1/4})$ for the FPU-$\beta$ model and $T_{eq}\sim \exp(-\varepsilon^{1/8})$ for the FPU-$\alpha\beta$ model, respectively, to the power-law $T_{eq}\sim \varepsilon^{-9/4}$ for both cases~\cite{Benettin2011, Benettin2013}.

Recently, the power-law relationship is affirmed~\cite{Onorato4208, PhysRevLett.120.144301, 0295-5075-121-4-44003, arXiv:1810.06902} based on the wave turbulence (WT) theory~\cite{1992kstbookZ, Majda1997, ZAKHAROV2001573, ZAKHAROV20041, 2011LNP825N, 2018htdbookZ}. It was shown analytically that the exact nontrivial six-wave resonant interactions are responsible for thermalization of short FPU chains in the weak nonlinear regime and, consequently, lead to $T_{eq}\sim \gamma^{-8}$ for the FPU-$\alpha$~\cite{Onorato4208} and $T_{eq}\sim \gamma^{-4}$ for the FPU-$\beta$~\cite{PhysRevLett.120.144301} model, where $\gamma$ is the nonlinearity strength defined, respectively, as $\gamma=\alpha \varepsilon^{1/2}$ and $\gamma=\beta \varepsilon$ for the two models ($\alpha$ and $\beta$ are the coefficient, respectively, of the cubic and quartic nonlinear term).~These results imply that any weak nonlinearity can ensure the system to be thermalized eventually. It was further conjectured (but not verified) that the nontrivial four-wave resonant interactions would dominate the thermalization process in the thermodynamic limit, leading to $T_{eq}\sim \gamma^{-4}$ and $T_{eq}\sim \gamma^{-2}$ for the FPU-$\alpha$~\cite{Onorato4208} and FPU-$\beta$~\cite{PhysRevLett.120.144301} model, respectively.

These conjectures were partially verified in a very recent effort~\cite{Our2018} where it was found that in the thermodynamic limit, a universal law, $T_{eq}\sim \gamma^{-2}$, applies generally to a class of one-dimensional (1D) lattices with interaction potential $V(x)=x^2/2+\lambda x^n/n$, where $n\geq4$ is an integer and $\gamma= \lambda \varepsilon^{(n-2)/2}$ is the nonlinearity strength. It also applies to another class of 1D lattices with symmetric interaction potential $V(x)=x^2/2+\lambda|x|^d/d$, where $d=m_1/m_2>2$ with $m_1$ and $m_2$ being two coprime integers and the nonlinearity strength $\gamma= \lambda \varepsilon^{(d-2)/2}$. The existence of this universal law strongly confirms the assumption that the exact nontrivial wave-wave resonances dominate thermalization.~However, it was also found that for a lattice with asymmetric potential interaction, though $T_{eq}$ still depends on $\gamma$ in a power law, the exponent deviates from $-2$.~In addition, the numerical result $T_{eq}\sim \gamma^{-4.6}$ for the asymmetric FPU-$\alpha$ model~\cite{Our2018} deviates from the conjectured $T_{eq}\sim \gamma^{-4}$~\cite{Onorato4208} seriously as well.

Note that in all these works, the studied nonlinear model was considered to be a perturbed harmonic lattice. The harmonic lattice is integrable and linear. However, for a given nonlinear model, it can also be viewed alternatively as a perturbed Toda lattice~\cite{1967431} that is integrable but nonlinear.~Interestingly, for some models, taking the latter viewpoint has been shown to be more consistent.~The FPU-$\alpha$ model is a good example, for which supporting evidence from various aspects, e.g., by a normal mode approach~\cite{FERGUSON1982157}, by the Lyapunov exponent analysis~\cite{PhysRevE.55.6566, PhysRevE.62.6078, Benettin2018}, and by thermalization process comparison~\cite{Benettin2011, Benettin2013}, has been found. In particular, as inspired by Refs.~\cite{Benettin2011, Benettin2013}, we have revisited the previous studies of thermalization and found that they strongly suggest the FPU-$\alpha$ model (the case of $n=3$ in Ref.~\cite{Our2018}) be viewed as the perturbed Toda lattice while other models as perturbed harmonic lattices. Given this, we were led to the conclusion that $T_{eq}\sim \gamma^{-2}$ generally applies to the perturbed harmonic lattices in the thermodynamic limit but not to the systems out of this class~\cite{Our2018}.

In the present work, we study systematically the thermalization rate of 1D perturbed Toda lattices in order to find whether there exists a universal law of $T_{eq}$ for this class as well and if the answer is yes, how it differs from $T_{eq}\sim \gamma^{-2}$ applicable to the 1D perturbed harmonic lattices. In the following, we will first introduce the models in the next section, then provide our theoretical arguments in Sec.~\ref{sec:3}. The numerical approach, as well as simulation results, will be described and presented in Sec.~\ref{sec:4}, followed by the summary and discussions in Sec.~\ref{sec:5}.

\section{\label{sec:2}The Models}

We study the perturbed Toda models with potential
\begin{equation}\label{eqVPT}
  V(x)=V_T(\alpha,x)+{\theta_nx^n}/{n}, \text{and}~n\geq3,
\end{equation}
where
\begin{equation}\label{eqVT}
  V_T(\alpha,x)=\frac{e^{2\alpha x}-2\alpha x-1}{4\alpha^2}
\end{equation}
is the Toda potential~\cite{1967431} with $\alpha$ being a free parameter and  $\theta_n$ is the coefficient of the perturbation. The Toda potential can be expanded as Taylor's series:
\begin{equation}\label{eqVTs}
  V_T(\alpha,x)=\frac{x^2}{2}+\frac{\alpha x^3}{3} +\sum_{n=4}^\infty\frac{\theta_n^Tx^n}{n},
\end{equation}
with the coefficients
\begin{equation}\label{eqTs}
  \theta_n^T=\frac{(2\alpha)^{n-2}}{(n-1)!}.
\end{equation}
From Eq.~(\ref{eqVTs}), we can see that the harmonic model is a special case of the Toda model when $\alpha=0$. For this reason, any nonlinear model can be regarded as a perturbed Toda model as well.

To make our analysis more generalisable, we also study the generalized FPU model~\cite{Benettin2011} with potential
\begin{equation}\label{eqVg}
  V_\text{gFPU}(x)=\frac{x^2}{2}+\frac{\alpha x^3}{3} +\sum_{n=4}^\infty\frac{\theta_nx^n}{n},
\end{equation}
where $\alpha$ and $\theta_n$ are free parameters. In principle, any smooth nonlinear potential can be written in this form.~Comparing with the perturbed Toda model given by Eq.~(\ref{eqVPT}) whose potential is perturbed only in a single high-order term, the potential of this model can be perturbed in multiple high-order terms.

\section{\label{sec:3}Definition of perturbation strength and theoretical analysis}

The Hamiltonian of our systems can be written as
\begin{equation}\label{eqHh0}
  H=H_0+H',
\end{equation}
where $H_0$ and $H'$ denote, respectively, the integrable part and the perturbation. Intuitively, the larger the perturbation is, the easier the system will be thermalized.~A conventional practice is to take the Hamiltonian of the harmonic lattice as $H_0$ and defines the rest nonlinear part as the perturbation.~However, the nonlinearity may not always be a good indicator to characterize the equipartition time.~For example, the Toda model can own a very strong nonlinearity but will never be thermalized due to its integrability.~Therefore, it is more reasonable to define perturbation strength as the $nonintegrability$, instead.~To this end, it would be superior to adopt the Hamiltonian of the Toda model as $H_0$.~This scenario is general; it also covers the conventional one where $H_0$ is the Hamiltonian of the harmonic lattice when $\alpha=0$. The definition of perturbation strength for different cases is given below:

{\it{The perturbed Toda model with $n\geq4$.}}---We get the perturbation by comparing Eq.~(\ref{eqVPT}) and Eq.~(\ref{eqVTs}) as
\begin{equation}\label{eqDis}
  H'=\frac{\theta_nx^n}{n}.
\end{equation}
By normalizing the Hamiltonian, i.e., by rescaling the relative displacement with the energy density so that $x'=\varepsilon^{1/2}x$, we can obtain the dimensionless perturbation strength as
\begin{equation}\label{eq:Epn}
\epsilon_n=|\theta_n|\varepsilon^{(n-2)/2}.
\end{equation}

{\it{The perturbed Toda model with $n=3$.}}---For this case, Eq.~(\ref{eqVPT}) can be rewritten in Taylor's series as
\begin{equation}\label{eq:VTPP}
  V(x)=\frac{x^2}{2}+\frac{(\alpha+\theta_3)x^3}{3}+\frac{\alpha^2x^4}{6}+\frac{\alpha^3x^5}{15}+\cdots.
\end{equation}
To set $\tilde{\alpha}=\alpha+\theta_3$ and with the help of Eq.~(\ref{eqTs}), we can get another Toda potential much closer to the perturbed system than the original one:
\begin{equation}\label{eq:VTPPP}
  V_{T}(\tilde{\alpha},x)=\frac{x^2}{2}+\frac{\tilde{\alpha}x^3}{3}+\frac{\tilde{\alpha}^2x^4}{6}+\frac{\tilde{\alpha}^3x^5}{15}+\cdots.
\end{equation}
Comparing Eq.~(\ref{eq:VTPP}) with Eq.~(\ref{eq:VTPPP}), we get perturbation
\begin{equation}
  H'=\frac{\epsilon_4x^4}{6}+\frac{\epsilon_5x^5}{15}+\cdots,
\end{equation}
where
\begin{equation}\label{eq:VTX3}
  \epsilon_n= |(\alpha+\theta_3)^{n-2}-\alpha^{n-2}|\varepsilon^{(n-2)/2},~n=4,5,\cdots,
\end{equation}
denotes the $n$th-order perturbation strength. Note that the leading perturbation is still of the 4th-order for $n=3$. Comparing with $\epsilon_n\sim |\theta_n|$ for the case of $n\geq4$ at a fixed $\varepsilon$, here $\epsilon_n$ has a more complicated relationship with $\theta_3$.

{\it{The generalized FPU model.}}~---~Similarly, comparing Eq.~(\ref{eqVg}) with Eq.~(\ref{eqVTs}), the perturbation can be identified to be
\begin{equation}
  H'=\sum_{n=4}^\infty\epsilon_n\frac{x^n}{n},
\end{equation}
where $\epsilon_n$ is the dimensionless strength of the $n$th-order perturbation, given by
\begin{equation}\label{eq:EpnFPU}
\epsilon_n= |\theta_n-\theta_n^T|\varepsilon^{(n-2)/2},~n=4,5,\cdots,\infty.
\end{equation}
Again, for this case the 4th-order perturbation is the lowest order one. For the FPU-$\alpha$ model, the leading perturbation strength is $\epsilon_4=\frac{2}{3}\alpha^2$, which is very different from the nonlinear strength $\alpha$ with respect to the linear integrable (harmonic) model.

Now let us evaluate the equipartition time.~Based on to the WT theory, it has been proved that either there are no resonances, or all of the scattering matrices are zero at all orders on the resonant manifold for integrable systems~\cite{ZAKHAROV1988283} so that they are characterized by trivial scattering processes and are never thermalized~\cite{Onorato4208}. All the scattering matrix being zero is broken when the integrable system is perturbed. We assume that the exact nontrivial $n$-wave scattering processes caused by the $n$th-order perturbation, in the thermodynamic limit, dominate the thermalization process of the perturbed system.~Then based on the theoretical results derived from the WT theory~\cite{Onorato4208, PhysRevLett.120.144301, 0295-5075-121-4-44003, Our2018}, the time scale of equipartition is
\begin{equation}\label{eqTeqn}
  T_{eq}\propto \epsilon_{n_L}^{-2},
\end{equation}
where $\epsilon_{n_L}$ and $n_L$ denote, respectively, the strength and the order of the leading perturbation. Hereafter we will show that the above assumption is supported by extensive numerical simulations.

\section{\label{sec:4}Numerical method and results}

For a homogeneous lattice we consider here that consists of $N+1$ particles of unit mass, labelled $0$, $1$, $2$, $\cdots$, $N$ from the left to the right, its Hamiltonian is
\begin{equation}\label{eqHam}
  H=\sum_{j=1}^N\left[\frac{p_j^2}{2}+V(q_{j}-q_{j-1})\right],
\end{equation}
where $p_j$ and $q_j$ are, respectively, the momentum and the displacement from the equilibrium position of the $j$th particle, and $V$ is the nearest-neighboring interaction potential.

\begin{figure*}[t]
  \centering
  \includegraphics[width=1.8\columnwidth]{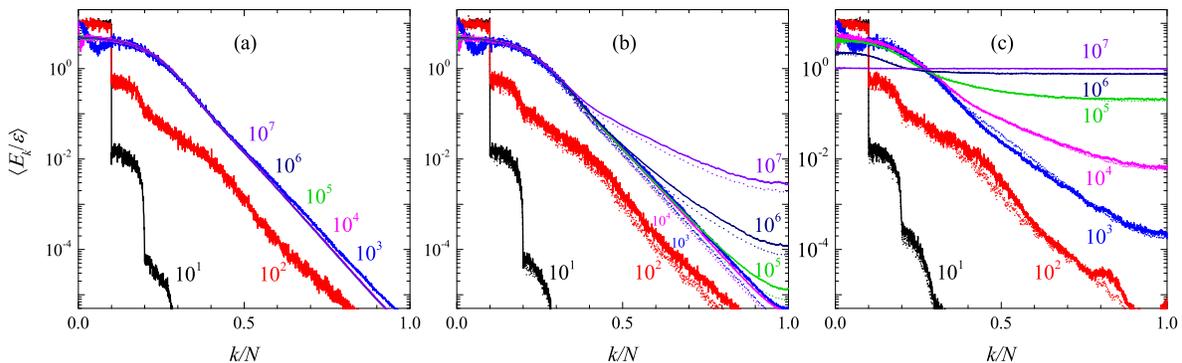}\\
  \caption{(a) The function $\langle E_k(t)/\varepsilon\rangle$ versus $k/N$ at various times for the Toda model. (b) and (c) show the results for the perturbed Toda model with $n=3$ and $n=4$, respectively. In (b), $\theta_3=-0.05$ (solid lines) and $0.05$ (dot lines), respectively; In (c), $\theta_4=-3$ (solid lines) and $3$ (dot lines), respectively. For all the cases $\varepsilon=10^{-3}$.}\label{fig:1}
\end{figure*}

For the fixed boundary conditions, i.e., $q_0=p_0=q_{N}=p_{N}=0$, the normal modes are defined as
\begin{align}
\begin{cases}
    Q_k&=\sqrt{\frac{2}{N}}\sum_{n=1}^Nq_n\sin\left(\frac{nk\pi}{N}\right),\\
    P_k&=\sqrt{\frac{2}{N}}\sum_{n=1}^Np_n\sin\left(\frac{nk\pi}{N}\right).
\end{cases}
\end{align}
To each mode $k$ one can associate a harmonic energy
\begin{equation}
    E_k=\frac{1}{2}\left(P_k^2+\omega_k^2Q_k^2\right)
\end{equation}
and a phase $\varphi_k$ defined via
\begin{equation}
   Q_k=\sqrt{2E_k/\omega_k^2}\sin{\left(\varphi_k\right)},~P_k=\sqrt{2E_k}\cos{\left(\varphi_k\right)}.
\end{equation}
Following the definition of equipartition, one expects
\begin{equation}
  \lim_{T\rightarrow\infty}\bar{E}_k(T)\simeq\varepsilon, \quad k=1,~\cdots,~N,
\end{equation}
where $\varepsilon=E/(N-1)$ is the energy density ($E$ denotes the total energy of the system) and $\bar{E}_k(T)$ represents the time average of $E_k$ up to time $T$; i.e.,
\begin{equation}\label{eq:EkT}
  \bar{E}_k(T)=\frac{1}{(1-\mu)T}\int_{\mu T}^TE_k(P(t),Q(t))dt.
\end{equation}
Here $\mu\in[0,1)$ controls the size of time average window. In our numerical simulations, $\mu=2/3$ is fixed, which not only can speed up the calculations, but also has the advantage of a quicker loss of the memory of the very special initial state as proposed in Ref.~\cite{Benettin2011}.

Based on the defined $\bar{E}_k(T)$, we need introduce a parameter to measure how close the system is to equipartition. A frequently used parameter is the effective relative number of degrees of freedom~\cite{PhysRevA.31.1039, GOEDDE1992200}. Here we employ the quantity $\xi(t)$ as in Ref. \cite{Benettin2011}, i.e.,
\begin{equation}\label{eq:xi}
  \xi(t)=\tilde{\xi}(t)\frac{e^{\eta(t)}}{N/2},
\end{equation}
where
\begin{equation}\label{eq:eta}
  \eta(t)=-\sum_{k=N/2}^{N}w_k(t)\log[w_k(t)]
\end{equation}
is the spectral entropy and
\begin{equation}
  \tilde{\xi}(t)=\frac{\sum_{k=N/2}^N\bar{E}_k(t)}{\frac{1}{2}\sum_{1\leq{k}\leq{N}}\bar{E}_k(t)},~
  w_k(t)=\frac{\bar{E}_k(t)}{\sum_{j=N/2}^N\bar{E}_j(t)}.
\end{equation}
When equipartition is approached, $\xi$ will saturate at $1$.

To integrate the motion equations numerically, we take the eighth-order Yoshida method~\cite{YOSHIDA1990262}. The typical time step is $\Delta t=0.1$; the corresponding relative error in energy conservation, when all modes are excited and do contribute to the total energy, is around $10^{-5}$. A further decrease of the time step by one order of magnitude, i.e., $\Delta t=0.01$, does not change the results. To suppress fluctuations, the average is done over $24$ phases uniformly distributed in $[0,2\pi]$, and we use $\langle\cdot\rangle$ to denote the ensemble average results. Initially the lowest $10\%$ of frequency modes are excited, $\alpha=-1$ and $N=2048$ are kept fixed throughout for all the numerical results presented. We have checked and verified that no qualitative difference will be resulted in neither when the percentage of the excited modes is changed nor when the system size is increased further.

In Fig.~\ref{fig:1}(a), the results of $\langle E_k(t)/\varepsilon\rangle$ versus $k/N$ for the Toda model are presented. It can be seen that only a small portion of the energy spread quickly from the
initial excited low-frequency modes to the high-frequency modes, then the energy profile keeps a stable localized form with an exponential decaying tail. It suggests that for the Toda model, the thermalized state can never be reached. In Fig.~\ref{fig:1}(b), the results for the perturbed Toda model with $\theta_3=-0.05$, $\epsilon_4=1.025\times10^{-4}$ (solid lines) and $\theta_3=0.05$, $\epsilon_4=0.975\times10^{-4}$ (dotted lines) are plotted. It can be seen that thermalization is faster approached in the former case as a consequence of the tiny difference of $\epsilon_4$, suggesting that the thermalization rate depends on the perturbation strength sensitively. It can also be seen that the energy of low-frequency modes remains for a long time and the energy of high-frequency modes increases very slowly, known as a signature of the metastable state~\cite{Benettin2009, Benettin2011, doi:10.1063/1.3658620, Benettin2013, arXiv:1810.06121}. Nevertheless, the system will be thermalized eventually. As a comparison, Fig.~\ref{fig:1}(c) shows the results of the perturbed Toda model with $\theta_4=-3$, $\epsilon_4=3\times10^{-3}$ (solid lines) and $\theta_4=3$, $\epsilon_4=3\times10^{-3}$ (dotted lines). Note that the thermalization rates of the two cases keep the same as the perturbation strengthes are identical. In addition, the system is fully thermalized at time $T\sim 10^7$, when $\langle E_k/\varepsilon\rangle=1$.

\begin{figure}[t]
  \centering
  \includegraphics[width=1\columnwidth]{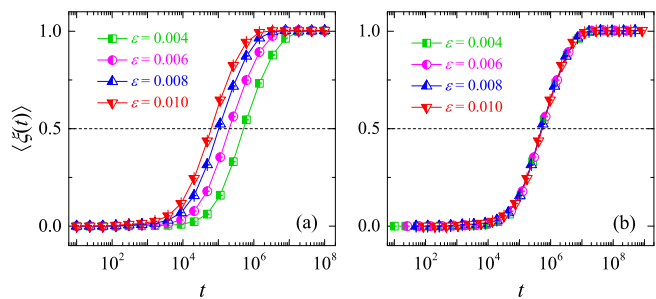}\\
  \caption{(a) The function $\langle\xi(t)\rangle$ for the perturbed Toda model with $\theta_3=0.5$ and different energy density $\varepsilon$ in the semi-log scale. (b) The same as (a) but the curves are shifted properly in the horizontal direction (with that for $\varepsilon=0.004$ unshifted) so that they overlap with each other perfectly.}\label{fig:2}
\end{figure}

To obtain the equipartition time, we study the properties of $\langle \xi(t)\rangle$ defined by Eq.~(\ref{eq:xi}).~Figure~\ref{fig:2}(a) shows the results for the perturbed Toda model with $\theta_3=0.5$. By varying energy density, $\varepsilon$, $\epsilon_n$ is changed. Note that on a sufficiently large time scale, all values of $\langle \xi(t)\rangle$ increase from $0$ to $1$ with very similar sigmoidal profiles. It suggests that energy equipartition is finally achieved. Meanwhile, when the energy density decreases, the time required to reach the thermalized state increases. Now we adopt the definition of the equipartition time, $T_{eq}$, as that when $\langle \xi(t)\rangle$ reaches the threshold value $0.5$ as in Refs.~\cite{Benettin2011, PhysRevE.93.022216}. Though assuming the threshold value $0.5$ is artificial, it does not influence the scaling law of $T_{eq}$~\cite{2008LNP728G}. This can be seen from Fig.~\ref{fig:2}(b), where the sigmoidal profiles in Fig.~\ref{fig:2}(a) can overlap with each other upon suitable shifts, which suggests that the concrete threshold value does not affect the scaling exponent of $T_{eq}$. With these preparations, we are ready to present the results of $T_{eq}$ as a function of $\epsilon_n$.

\begin{figure}[t]
  \centering
  \includegraphics[width=1\columnwidth]{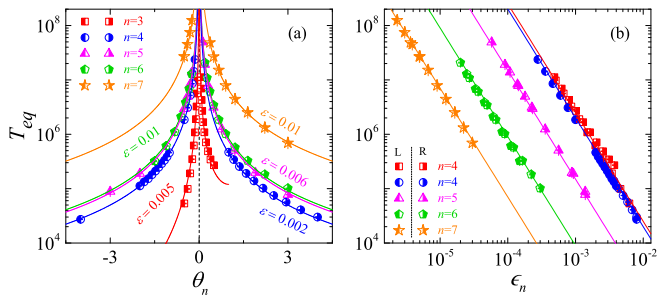}\\
  \caption{(a) The thermalization time $T_{eq}$ of the perturbed Toda model as a function of $\theta_n$ for $n=3$, $4$, $5$, $6$, and $7$ in semi-log scale.~The solid lines of $\Lambda$-shape are for $T_{eq}\sim \epsilon_n^{-2}$ with the best fitting prefactors, which are plotted for reference. Here $\epsilon_n$ is defined by Eq.~(\ref{eq:VTX3}) for $n=3$ and by Eq.~(\ref{eq:Epn}) for $n=4, 5, 6$, and $7$.~(b) The same as (a) but plotted as a function of $\epsilon_n$ in log-log scale instead. Solid lines with slope $-2$ are drawn for reference.~The letters L and R in the legend indicate the points to the left and right of the peak [see (a)], respectively.}\label{fig:TeqPT}
\end{figure}

In Fig.~\ref{fig:TeqPT}(a), the numerical results of $T_{eq}$ as a function of $\theta_{3}$, $\theta_{4}$, $\theta_{5}$, $\theta_{6}$, and $\theta_{7}$ are shown in semi-log scale for the perturbed Toda model. As for a given energy density, $T_{eq}$ for different $n$ could be remarkably distinct, here we adopt different energy density for different $n$ in order to present all the data in a single picture. We can see that all the numerical points can be well fitted with a $\Lambda$-shape curve of form $T_{eq}\sim |\theta_n|^{-2}$ for $n\geq4$ and $T_{eq}\sim |(\theta_3-1)^2-1|^{-2}$ for $n=3$, respectively, which is exactly what predicted by Eq.~(\ref{eqTeqn}) when Eq.~(\ref{eq:VTX3}) and Eq.~(\ref{eq:Epn}) for the two cases, respectively, are substituted into. The data of $T_{eq}$ are also plotted versus $\epsilon_n$ in Fig.~\ref{fig:TeqPT}(b) in log-log scale. Note that all the points fall on the lines with a slope of $-2$, suggesting $T_{eq}\sim \epsilon_n^{-2}$ holds for all the cases. As shown by Eq.~(\ref{eq:VTX3}), the relationship between $\epsilon_n$ and $\theta_3$ is very complicated. Next, we will carefully check if this relationship is true in a wider parameter range.

\begin{figure}[t]
  \centering
  \includegraphics[width=1\columnwidth]{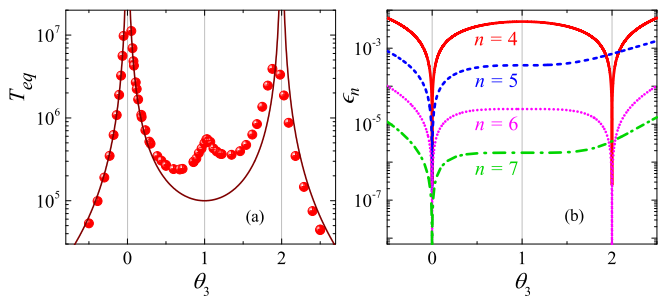}\\
  \caption{(a) The thermalization time $T_{eq}$ of the perturbed Toda model as a function of $\theta_3$ with fixed $\varepsilon=0.005$ in semi-log scale. The wine solid line of form $T_{eq}\sim \epsilon_4^{-2}$ is drawn for reference. (b) The dependence of $\epsilon_n$ on $\theta_3$ for $n=4, 5, 6$, and $7$ in semi-log scale.}\label{fig:TeqPTX3}
\end{figure}

Figure~\ref{fig:TeqPTX3}(a) shows $T_{eq}$ versus $\theta_3$ in a large parameter range, where three peaks of $T_{eq}$ can be clearly recognized. In order to understand the underlying mechanism for the formation of these peaks, we plot $\epsilon_n$ as a function of $\theta_3$ [see Eq.~(\ref{eq:VTX3})] in semi-log scale in Fig.~\ref{fig:TeqPTX3}(b). The first peak resides at $\theta_3=0$,  corresponding to the integrable point of the Toda lattice, i.e., $\epsilon_n=0$ for all $n$. The theoretical prediction gives $T_{eq}\sim \epsilon_4^{-2}\sim|(\theta_3-1)^2-1|^{-2}$ near this point, which agrees very well with the numerical results [see Fig.~\ref{fig:TeqPTX3}(a)]. It is amazing to realize that this theoretical result also predicts the location and the height of the third peak near $\theta_3=2=-2\alpha$, which corresponds to the special point of $\epsilon_n=0$ for all even $n$ [see Fig.~\ref{fig:TeqPTX3}(b)]. Near this point the perturbed Toda model is very close to its `mirror image' that adopts a minus $\alpha$ in the Toda potential [see Eq.~(\ref{eqVT})]. In particular, for $\theta_3=2$, the perturbed system can be regarded to have a mirrored Toda potential with additional the fifth and higher odd order perturbations [see blue dashed line and green dashed dots line in Fig.~\ref{fig:TeqPTX3}(b)]. Note that the third peak is not located at $\theta_3=2$ exactly due to the influence of high order resonances becoming non-negligible near this point. The middle peak is near $\theta_3=1=-\alpha$, where the cubic coefficient is approximately zero. Namely, the perturbed system is close to the linear integrable point (harmonic lattice) and hence has an approximately symmetric potential, such that the additional energy mixing channel introduced by the asymmetry of interaction potential~\cite{Our2018} is closed and thus gives rise to the middle peak.

So far our investigation has suggested that thermalization of weakly perturbed Toda lattices follows a common feature, i.e., $T_{eq}\propto\epsilon^{-2}$. This general behavior coincides completely with that for the perturbed harmonic lattices~\cite{Our2018}. The validity of our theoretical analysis can be further tested in the generalized FPU models, and meanwhile we find that it also gives satisfactory explanations to the numerical results reported previously~\cite{Benettin2011}.

\begin{figure*}[t]
  \centering
  \includegraphics[width=2\columnwidth]{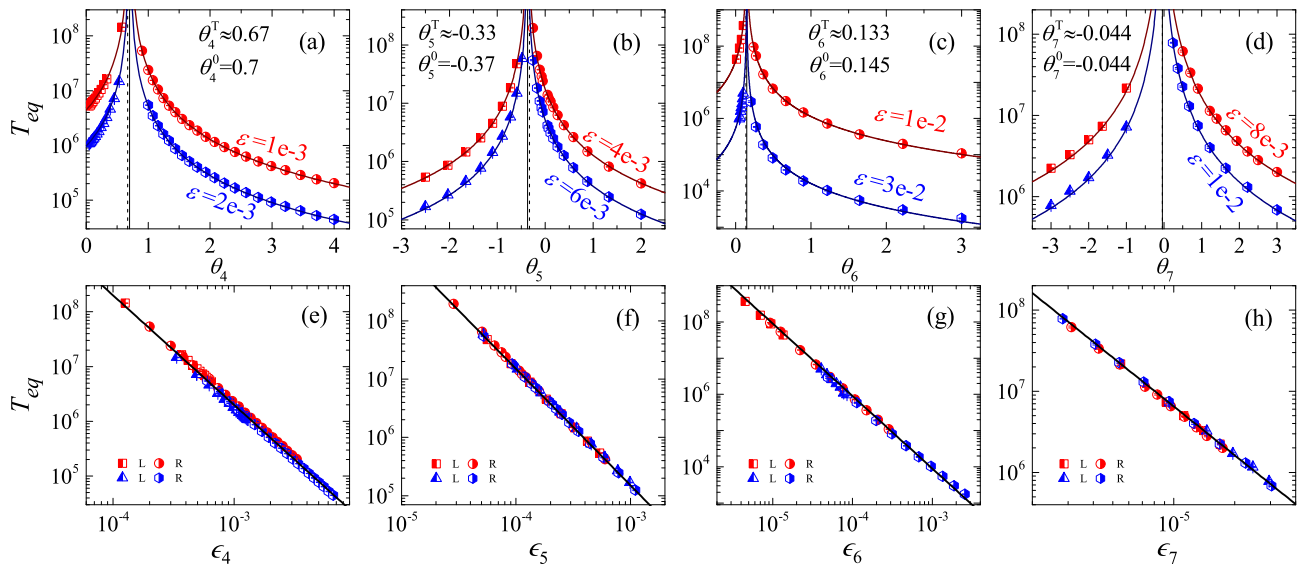}\\
  \caption{The thermalization time $T_{eq}$ as function of $\theta_4$ with $\theta_{5},\cdots,\theta_{\infty}=0$ (a); of $\theta_5$ with $\theta_4=\theta_4^T$ and $\theta_{6},\cdots,\theta_{\infty}=0$ (b); of $\theta_6$ with $\theta_n=\theta_n^T$~$(n=4,5)$ and $\theta_{7},\cdots,\theta_{\infty}=0$ (c); and of $\theta_7$ with $\theta_n=\theta_n^T~(n=4,5,6)$ and $\theta_{8},\cdots,\theta_{\infty}=0$ (d), in semi-log scale, respectively. The vertical dashed (solid) lines are for $\theta=\theta_n^T$ ($\theta=\theta_n^0$ ), and the $\Lambda$-shape solid lines are for $T_{eq}\sim |\theta_n-\theta_n^0|^{-2}$, which are plotted for reference. Panels (e)-(h): the same as in (a)-(d) but shown as a function of $\epsilon_n$ in log-log scale instead. The solid lines with slope $-2$ are drawn for reference. The letters L and R in the legend indicate the points to the left and right of the peak in (a)-(d), respectively.
  }\label{fig:TeqGFPU}
\end{figure*}

Figure~\ref{fig:TeqGFPU} summarizes the numerical results of the generalized FPU model.~Figures~\ref{fig:TeqGFPU}(a)-(d) show $T_{eq}$ as a function of $\theta_{4}$, $\theta_{5}$, $\theta_{6}$, and $\theta_{7}$ in semi-log scale, at two different energy densities, respectively. In Fig.~\ref{fig:TeqGFPU}(a) and (e), the results for repeating the previous study in Ref.~\cite{Benettin2011} are presented, and extended study results are presented in Figs.~\ref{fig:TeqGFPU}(b)-(d) and (f)-(h). From (a) to (d), each figure shows a very marked peak near $\theta_n^T$, and all the numerical points are well fitted with $T_{eq}\sim |\theta_n-\theta_n^0|^{-2}$. But the center of the peak is not exactly at $\theta_n^T$, which is mainly because the effect of higher order resonances becomes increasingly significant as $\theta_n$ tends to $\theta_n^T$. However, with the increase of $n$ the difference between $\theta_n^0$ and $\theta_n^T$ decreases due to the fact that the higher the order, the weaker its effect [see the values indicated in Figs.~\ref{fig:TeqGFPU}(a)-(d)]. In Figs.~\ref{fig:TeqGFPU}(e)-(h), $T_{eq}$ is redrawn as a function of $\epsilon_n$ that is defined by Eq.~(\ref{eq:EpnFPU}) with $\theta_n^T$ being replaced by $\theta_n^0$, considering the higher order correction. Note that all the data points fall onto the lines with slope $-2$, suggesting that again, $T_{eq}\sim \epsilon_n^{-2}$ is confirmed convincingly.

Notice that expression (\ref{eqTeqn}) can converge to the theoretical results of Ref.~\cite{Our2018}, i.e., $T_{eq}\propto\lambda^{-2}\varepsilon^{-(n-2)}$ for $n\geq4$, which can be regarded as a perturbed linear integrable systems, i.e., a special case of $\alpha=0$ in our study here. What is interesting is that the FPU-$\alpha$ model with $\alpha\neq0$ and $\theta_n=0$ is covered by expression (\ref{eqTeqn}) automatically. In this case, from Eq.~(\ref{eqTeqn}) we have $T_{eq}\propto \epsilon_4^{-2}=|0-\theta_4^T|^{-2}\varepsilon^{-2}=4/9\alpha^{-4}\varepsilon^{-2}$, which is the same as that given in Refs.~\cite{Onorato4208, Our2018}.

\section{\label{sec:5}Summary and Discussions}

In this work, we have shown that thermalization of a 1D weak nonlinear lattice exhibits a universal feature in the thermodynamic limit, i.e., $T_{eq}\propto\epsilon^{-2}$, where $\epsilon$ is the perturbation strength defined as the difference in the potential of the system from the Toda potential. This universal behavior supports the assumption within the WT framework that the exact nontrivial wave-wave resonances dominate the thermalization process of a weak nonlinear lattice. The key to identify the universal exponent $-2$ is to select the Toda lattice as the reference integrable system. In doing so, the third order nonlinearity has been found to be so crucial that it governs how we should assign the reference integrable system consistently.~In particular, the system with (without) the cubic term of interactions should be regarded to be the perturbed Toda (harmonic) model.~Comparing with previous studies, the resultant thermalization law ($T_{eq}\propto\epsilon^{-2}$) provides a unified and consistent picture for thermalization of one-dimensional nonlinear chains.

\section*{Acknowledgment}

We are indebted to Prof. Jiao Wang for his kind help in preparing the manuscript.~This work is supported by NSFC (Grant No. 11335006).

\bibliography{TodaReference}

\begin{thebibliography}{60}%
\makeatletter
\providecommand \@ifxundefined [1]{%
 \@ifx{#1\undefined}
}%
\providecommand \@ifnum [1]{%
 \ifnum #1\expandafter \@firstoftwo
 \else \expandafter \@secondoftwo
 \fi
}%
\providecommand \@ifx [1]{%
 \ifx #1\expandafter \@firstoftwo
 \else \expandafter \@secondoftwo
 \fi
}%
\providecommand \natexlab [1]{#1}%
\providecommand \enquote  [1]{``#1''}%
\providecommand \bibnamefont  [1]{#1}%
\providecommand \bibfnamefont [1]{#1}%
\providecommand \citenamefont [1]{#1}%
\providecommand \href@noop [0]{\@secondoftwo}%
\providecommand \href [0]{\begingroup \@sanitize@url \@href}%
\providecommand \@href[1]{\@@startlink{#1}\@@href}%
\providecommand \@@href[1]{\endgroup#1\@@endlink}%
\providecommand \@sanitize@url [0]{\catcode `\\12\catcode `\$12\catcode
  `\&12\catcode `\#12\catcode `\^12\catcode `\_12\catcode `\%12\relax}%
\providecommand \@@startlink[1]{}%
\providecommand \@@endlink[0]{}%
\providecommand \url  [0]{\begingroup\@sanitize@url \@url }%
\providecommand \@url [1]{\endgroup\@href {#1}{\urlprefix }}%
\providecommand \urlprefix  [0]{URL }%
\providecommand \Eprint [0]{\href }%
\providecommand \doibase [0]{http://dx.doi.org/}%
\providecommand \selectlanguage [0]{\@gobble}%
\providecommand \bibinfo  [0]{\@secondoftwo}%
\providecommand \bibfield  [0]{\@secondoftwo}%
\providecommand \translation [1]{[#1]}%
\providecommand \BibitemOpen [0]{}%
\providecommand \bibitemStop [0]{}%
\providecommand \bibitemNoStop [0]{.\EOS\space}%
\providecommand \EOS [0]{\spacefactor3000\relax}%
\providecommand \BibitemShut  [1]{\csname bibitem#1\endcsname}%
\let\auto@bib@innerbib\@empty
\bibitem [{\citenamefont {Fermi}\ \emph {et~al.}(1955)\citenamefont {Fermi},
  \citenamefont {Pasta}, \citenamefont {Ulam},\ and\ \citenamefont
  {Tsingou}}]{Fermi1955}%
  \BibitemOpen
  \bibfield  {author} {\bibinfo {author} {\bibfnamefont {E.}~\bibnamefont
  {Fermi}}, \bibinfo {author} {\bibfnamefont {P.}~\bibnamefont {Pasta}},
  \bibinfo {author} {\bibfnamefont {S.}~\bibnamefont {Ulam}}, \ and\ \bibinfo
  {author} {\bibfnamefont {M.}~\bibnamefont {Tsingou}},\ }\href@noop {}
  {\bibfield  {journal} {\bibinfo  {journal} {{Los Alamos Scientific
  Laboratory, Report No. LA-1940}}\ } (\bibinfo {year} {1955})}\BibitemShut
  {NoStop}%
\bibitem [{\citenamefont {Dauxois}(2008)}]{dauxois:ensl-00202296}%
  \BibitemOpen
  \bibfield  {author} {\bibinfo {author} {\bibfnamefont {T.}~\bibnamefont
  {Dauxois}},\ }\href {\doibase 10.1063/1.2835154} {\bibfield  {journal}
  {\bibinfo  {journal} {{Phys. Today}}\ ,\ \bibinfo {pages} {55}} (\bibinfo
  {year} {2008})}\BibitemShut {NoStop}%
\bibitem [{\citenamefont {Wu}\ and\ \citenamefont
  {Patton}(2007)}]{PhysRevLett.98.047202}%
  \BibitemOpen
  \bibfield  {author} {\bibinfo {author} {\bibfnamefont {M.}~\bibnamefont
  {Wu}}\ and\ \bibinfo {author} {\bibfnamefont {C.~E.}\ \bibnamefont
  {Patton}},\ }\href {\doibase 10.1103/PhysRevLett.98.047202} {\bibfield
  {journal} {\bibinfo  {journal} {Phys. Rev. Lett.}\ }\textbf {\bibinfo
  {volume} {98}},\ \bibinfo {pages} {047202} (\bibinfo {year}
  {2007})}\BibitemShut {NoStop}%
\bibitem [{\citenamefont {Mussot}\ \emph {et~al.}(2014)\citenamefont {Mussot},
  \citenamefont {Kudlinski}, \citenamefont {Droques}, \citenamefont
  {Szriftgiser},\ and\ \citenamefont {Akhmediev}}]{PhysRevX.4.011054}%
  \BibitemOpen
  \bibfield  {author} {\bibinfo {author} {\bibfnamefont {A.}~\bibnamefont
  {Mussot}}, \bibinfo {author} {\bibfnamefont {A.}~\bibnamefont {Kudlinski}},
  \bibinfo {author} {\bibfnamefont {M.}~\bibnamefont {Droques}}, \bibinfo
  {author} {\bibfnamefont {P.}~\bibnamefont {Szriftgiser}}, \ and\ \bibinfo
  {author} {\bibfnamefont {N.}~\bibnamefont {Akhmediev}},\ }\href {\doibase
  10.1103/PhysRevX.4.011054} {\bibfield  {journal} {\bibinfo  {journal} {Phys.
  Rev. X}\ }\textbf {\bibinfo {volume} {4}},\ \bibinfo {pages} {011054}
  (\bibinfo {year} {2014})}\BibitemShut {NoStop}%
\bibitem [{\citenamefont {Bao}\ \emph {et~al.}(2016)\citenamefont {Bao},
  \citenamefont {Jaramillo-Villegas}, \citenamefont {Xuan}, \citenamefont
  {Leaird}, \citenamefont {Qi},\ and\ \citenamefont
  {Weiner}}]{PhysRevLett.117.163901}%
  \BibitemOpen
  \bibfield  {author} {\bibinfo {author} {\bibfnamefont {C.}~\bibnamefont
  {Bao}}, \bibinfo {author} {\bibfnamefont {J.~A.}\ \bibnamefont
  {Jaramillo-Villegas}}, \bibinfo {author} {\bibfnamefont {Y.}~\bibnamefont
  {Xuan}}, \bibinfo {author} {\bibfnamefont {D.~E.}\ \bibnamefont {Leaird}},
  \bibinfo {author} {\bibfnamefont {M.}~\bibnamefont {Qi}}, \ and\ \bibinfo
  {author} {\bibfnamefont {A.~M.}\ \bibnamefont {Weiner}},\ }\href {\doibase
  10.1103/PhysRevLett.117.163901} {\bibfield  {journal} {\bibinfo  {journal}
  {Phys. Rev. Lett.}\ }\textbf {\bibinfo {volume} {117}},\ \bibinfo {pages}
  {163901} (\bibinfo {year} {2016})}\BibitemShut {NoStop}%
\bibitem [{\citenamefont {Guasoni}\ \emph {et~al.}(2017)\citenamefont
  {Guasoni}, \citenamefont {Garnier}, \citenamefont {Rumpf}, \citenamefont
  {Sugny}, \citenamefont {Fatome}, \citenamefont {Amrani}, \citenamefont
  {Millot},\ and\ \citenamefont {Picozzi}}]{PhysRevX.7.011025}%
  \BibitemOpen
  \bibfield  {author} {\bibinfo {author} {\bibfnamefont {M.}~\bibnamefont
  {Guasoni}}, \bibinfo {author} {\bibfnamefont {J.}~\bibnamefont {Garnier}},
  \bibinfo {author} {\bibfnamefont {B.}~\bibnamefont {Rumpf}}, \bibinfo
  {author} {\bibfnamefont {D.}~\bibnamefont {Sugny}}, \bibinfo {author}
  {\bibfnamefont {J.}~\bibnamefont {Fatome}}, \bibinfo {author} {\bibfnamefont
  {F.}~\bibnamefont {Amrani}}, \bibinfo {author} {\bibfnamefont
  {G.}~\bibnamefont {Millot}}, \ and\ \bibinfo {author} {\bibfnamefont
  {A.}~\bibnamefont {Picozzi}},\ }\href {\doibase 10.1103/PhysRevX.7.011025}
  {\bibfield  {journal} {\bibinfo  {journal} {Phys. Rev. X}\ }\textbf {\bibinfo
  {volume} {7}},\ \bibinfo {pages} {011025} (\bibinfo {year}
  {2017})}\BibitemShut {NoStop}%
\bibitem [{\citenamefont {Mussot}\ \emph {et~al.}(2018)\citenamefont {Mussot},
  \citenamefont {Naveau}, \citenamefont {Conforti}, \citenamefont {Kudlinski},
  \citenamefont {Copie}, \citenamefont {Szriftgiser},\ and\ \citenamefont
  {Trillo}}]{NaturePhotonics.12.303}%
  \BibitemOpen
  \bibfield  {author} {\bibinfo {author} {\bibfnamefont {A.}~\bibnamefont
  {Mussot}}, \bibinfo {author} {\bibfnamefont {C.}~\bibnamefont {Naveau}},
  \bibinfo {author} {\bibfnamefont {M.}~\bibnamefont {Conforti}}, \bibinfo
  {author} {\bibfnamefont {A.}~\bibnamefont {Kudlinski}}, \bibinfo {author}
  {\bibfnamefont {F.}~\bibnamefont {Copie}}, \bibinfo {author} {\bibfnamefont
  {P.}~\bibnamefont {Szriftgiser}}, \ and\ \bibinfo {author} {\bibfnamefont
  {S.}~\bibnamefont {Trillo}},\ }\href {\doibase 10.1038/s41566-018-0136-1}
  {\bibfield  {journal} {\bibinfo  {journal} {Nature Photonics}\ }\textbf
  {\bibinfo {volume} {12}},\ \bibinfo {pages} {303} (\bibinfo {year}
  {2018})}\BibitemShut {NoStop}%
\bibitem [{\citenamefont {Pierangeli}\ \emph {et~al.}(2018)\citenamefont
  {Pierangeli}, \citenamefont {Flammini}, \citenamefont {Zhang}, \citenamefont
  {Marcucci}, \citenamefont {Agranat}, \citenamefont {Grinevich}, \citenamefont
  {Santini}, \citenamefont {Conti},\ and\ \citenamefont
  {DelRe}}]{PhysRevX.8.041017}%
  \BibitemOpen
  \bibfield  {author} {\bibinfo {author} {\bibfnamefont {D.}~\bibnamefont
  {Pierangeli}}, \bibinfo {author} {\bibfnamefont {M.}~\bibnamefont
  {Flammini}}, \bibinfo {author} {\bibfnamefont {L.}~\bibnamefont {Zhang}},
  \bibinfo {author} {\bibfnamefont {G.}~\bibnamefont {Marcucci}}, \bibinfo
  {author} {\bibfnamefont {A.~J.}\ \bibnamefont {Agranat}}, \bibinfo {author}
  {\bibfnamefont {P.~G.}\ \bibnamefont {Grinevich}}, \bibinfo {author}
  {\bibfnamefont {P.~M.}\ \bibnamefont {Santini}}, \bibinfo {author}
  {\bibfnamefont {C.}~\bibnamefont {Conti}}, \ and\ \bibinfo {author}
  {\bibfnamefont {E.}~\bibnamefont {DelRe}},\ }\href {\doibase
  10.1103/PhysRevX.8.041017} {\bibfield  {journal} {\bibinfo  {journal} {Phys.
  Rev. X}\ }\textbf {\bibinfo {volume} {8}},\ \bibinfo {pages} {041017}
  (\bibinfo {year} {2018})}\BibitemShut {NoStop}%
\bibitem [{\citenamefont {Zabusky}\ and\ \citenamefont
  {Kruskal}(1965)}]{PhysRevLett.15.240}%
  \BibitemOpen
  \bibfield  {author} {\bibinfo {author} {\bibfnamefont {N.~J.}\ \bibnamefont
  {Zabusky}}\ and\ \bibinfo {author} {\bibfnamefont {M.~D.}\ \bibnamefont
  {Kruskal}},\ }\href {\doibase 10.1103/PhysRevLett.15.240} {\bibfield
  {journal} {\bibinfo  {journal} {Phys. Rev. Lett.}\ }\textbf {\bibinfo
  {volume} {15}},\ \bibinfo {pages} {240} (\bibinfo {year} {1965})}\BibitemShut
  {NoStop}%
\bibitem [{\citenamefont {Dauxois}\ and\ \citenamefont
  {Peyrard}(2006)}]{Dauxois:2006zz}%
  \BibitemOpen
  \bibfield  {author} {\bibinfo {author} {\bibfnamefont {T.}~\bibnamefont
  {Dauxois}}\ and\ \bibinfo {author} {\bibfnamefont {M.}~\bibnamefont
  {Peyrard}},\ }\href@noop {} {\emph {\bibinfo {title} {{Physics of
  solitons}}}}\ (\bibinfo  {publisher} {Cambridge University Press},\ \bibinfo
  {year} {2006})\BibitemShut {NoStop}%
\bibitem [{\citenamefont {Gardner}\ \emph {et~al.}(1967)\citenamefont
  {Gardner}, \citenamefont {Greene}, \citenamefont {Kruskal},\ and\
  \citenamefont {Miura}}]{PhysRevLett.19.1095}%
  \BibitemOpen
  \bibfield  {author} {\bibinfo {author} {\bibfnamefont {C.~S.}\ \bibnamefont
  {Gardner}}, \bibinfo {author} {\bibfnamefont {J.~M.}\ \bibnamefont {Greene}},
  \bibinfo {author} {\bibfnamefont {M.~D.}\ \bibnamefont {Kruskal}}, \ and\
  \bibinfo {author} {\bibfnamefont {R.~M.}\ \bibnamefont {Miura}},\ }\href
  {\doibase 10.1103/PhysRevLett.19.1095} {\bibfield  {journal} {\bibinfo
  {journal} {Phys. Rev. Lett.}\ }\textbf {\bibinfo {volume} {19}},\ \bibinfo
  {pages} {1095} (\bibinfo {year} {1967})}\BibitemShut {NoStop}%
\bibitem [{\citenamefont {{Izrailev}}\ and\ \citenamefont
  {{Chirikov}}(1966)}]{1966SPhD1130I}%
  \BibitemOpen
  \bibfield  {author} {\bibinfo {author} {\bibfnamefont {F.~M.}\ \bibnamefont
  {{Izrailev}}}\ and\ \bibinfo {author} {\bibfnamefont {B.~V.}\ \bibnamefont
  {{Chirikov}}},\ }\href {http://adsabs.harvard.edu/abs/1966SPhD...11...30I}
  {\bibfield  {journal} {\bibinfo  {journal} {Sov. Phys. Dokl.}\ }\textbf
  {\bibinfo {volume} {11}},\ \bibinfo {pages} {30} (\bibinfo {year}
  {1966})}\BibitemShut {NoStop}%
\bibitem [{\citenamefont {Chirikov}(1979)}]{CHIRIKOV1979263}%
  \BibitemOpen
  \bibfield  {author} {\bibinfo {author} {\bibfnamefont {B.~V.}\ \bibnamefont
  {Chirikov}},\ }\href {\doibase 10.1016/0370-1573(79)90023-1} {\bibfield
  {journal} {\bibinfo  {journal} {Phys. Rep.}\ }\textbf {\bibinfo {volume}
  {52}},\ \bibinfo {pages} {263 } (\bibinfo {year} {1979})}\BibitemShut
  {NoStop}%
\bibitem [{\citenamefont {Campbell}\ \emph {et~al.}(2005)\citenamefont
  {Campbell}, \citenamefont {Rosenau},\ and\ \citenamefont
  {Zaslavsky}}]{doi:10.1063/1.1889345}%
  \BibitemOpen
  \bibfield  {author} {\bibinfo {author} {\bibfnamefont {D.~K.}\ \bibnamefont
  {Campbell}}, \bibinfo {author} {\bibfnamefont {P.}~\bibnamefont {Rosenau}}, \
  and\ \bibinfo {author} {\bibfnamefont {G.~M.}\ \bibnamefont {Zaslavsky}},\
  }\href {\doibase 10.1063/1.1889345} {\bibfield  {journal} {\bibinfo
  {journal} {Chaos}\ }\textbf {\bibinfo {volume} {15}},\ \bibinfo {pages}
  {015101} (\bibinfo {year} {2005})}\BibitemShut {NoStop}%
\bibitem [{\citenamefont {Berman}\ and\ \citenamefont
  {Izrailev}(2005)}]{doi:10.1063/1.1855036}%
  \BibitemOpen
  \bibfield  {author} {\bibinfo {author} {\bibfnamefont {G.~P.}\ \bibnamefont
  {Berman}}\ and\ \bibinfo {author} {\bibfnamefont {F.~M.}\ \bibnamefont
  {Izrailev}},\ }\href {\doibase 10.1063/1.1855036} {\bibfield  {journal}
  {\bibinfo  {journal} {Chaos}\ }\textbf {\bibinfo {volume} {15}},\ \bibinfo
  {pages} {015104} (\bibinfo {year} {2005})}\BibitemShut {NoStop}%
\bibitem [{\citenamefont {Pettini}\ \emph {et~al.}(2005)\citenamefont
  {Pettini}, \citenamefont {Casetti}, \citenamefont {Cerruti-Sola},
  \citenamefont {Franzosi},\ and\ \citenamefont
  {Cohen}}]{doi:10.1063/1.1849131}%
  \BibitemOpen
  \bibfield  {author} {\bibinfo {author} {\bibfnamefont {M.}~\bibnamefont
  {Pettini}}, \bibinfo {author} {\bibfnamefont {L.}~\bibnamefont {Casetti}},
  \bibinfo {author} {\bibfnamefont {M.}~\bibnamefont {Cerruti-Sola}}, \bibinfo
  {author} {\bibfnamefont {R.}~\bibnamefont {Franzosi}}, \ and\ \bibinfo
  {author} {\bibfnamefont {E.~G.~D.}\ \bibnamefont {Cohen}},\ }\href {\doibase
  10.1063/1.1849131} {\bibfield  {journal} {\bibinfo  {journal} {Chaos}\
  }\textbf {\bibinfo {volume} {15}},\ \bibinfo {pages} {015106} (\bibinfo
  {year} {2005})}\BibitemShut {NoStop}%
\bibitem [{\citenamefont {Zabusky}(2005)}]{doi:10.1063/1.1861554}%
  \BibitemOpen
  \bibfield  {author} {\bibinfo {author} {\bibfnamefont {N.~J.}\ \bibnamefont
  {Zabusky}},\ }\href {\doibase 10.1063/1.1861554} {\bibfield  {journal}
  {\bibinfo  {journal} {Chaos}\ }\textbf {\bibinfo {volume} {15}},\ \bibinfo
  {pages} {015102} (\bibinfo {year} {2005})}\BibitemShut {NoStop}%
\bibitem [{\citenamefont {Porter}\ \emph {et~al.}(2009)\citenamefont {Porter},
  \citenamefont {Zabusky}, \citenamefont {Hu},\ and\ \citenamefont
  {Campbell}}]{Porter2009Fermi}%
  \BibitemOpen
  \bibfield  {author} {\bibinfo {author} {\bibfnamefont {M.~A.}\ \bibnamefont
  {Porter}}, \bibinfo {author} {\bibfnamefont {N.~J.}\ \bibnamefont {Zabusky}},
  \bibinfo {author} {\bibfnamefont {B.}~\bibnamefont {Hu}}, \ and\ \bibinfo
  {author} {\bibfnamefont {D.~K.}\ \bibnamefont {Campbell}},\ }\href {\doibase
  10.1511/2009.78.214} {\bibfield  {journal} {\bibinfo  {journal} {American
  Scientist}\ }\textbf {\bibinfo {volume} {97}},\ \bibinfo {pages} {214}
  (\bibinfo {year} {2009})}\BibitemShut {NoStop}%
\bibitem [{\citenamefont {Lepri}(1998)}]{PhysRevE.58.7165}%
  \BibitemOpen
  \bibfield  {author} {\bibinfo {author} {\bibfnamefont {S.}~\bibnamefont
  {Lepri}},\ }\href {\doibase 10.1103/PhysRevE.58.7165} {\bibfield  {journal}
  {\bibinfo  {journal} {Phys. Rev. E}\ }\textbf {\bibinfo {volume} {58}},\
  \bibinfo {pages} {7165} (\bibinfo {year} {1998})}\BibitemShut {NoStop}%
\bibitem [{\citenamefont {Flach}\ \emph {et~al.}(2005)\citenamefont {Flach},
  \citenamefont {Ivanchenko},\ and\ \citenamefont
  {Kanakov}}]{PhysRevLett.95.064102}%
  \BibitemOpen
  \bibfield  {author} {\bibinfo {author} {\bibfnamefont {S.}~\bibnamefont
  {Flach}}, \bibinfo {author} {\bibfnamefont {M.~V.}\ \bibnamefont
  {Ivanchenko}}, \ and\ \bibinfo {author} {\bibfnamefont {O.~I.}\ \bibnamefont
  {Kanakov}},\ }\href {\doibase 10.1103/PhysRevLett.95.064102} {\bibfield
  {journal} {\bibinfo  {journal} {Phys. Rev. Lett.}\ }\textbf {\bibinfo
  {volume} {95}},\ \bibinfo {pages} {064102} (\bibinfo {year}
  {2005})}\BibitemShut {NoStop}%
\bibitem [{\citenamefont {Flach}\ \emph {et~al.}(2006)\citenamefont {Flach},
  \citenamefont {Ivanchenko},\ and\ \citenamefont
  {Kanakov}}]{PhysRevE.73.036618}%
  \BibitemOpen
  \bibfield  {author} {\bibinfo {author} {\bibfnamefont {S.}~\bibnamefont
  {Flach}}, \bibinfo {author} {\bibfnamefont {M.~V.}\ \bibnamefont
  {Ivanchenko}}, \ and\ \bibinfo {author} {\bibfnamefont {O.~I.}\ \bibnamefont
  {Kanakov}},\ }\href {\doibase 10.1103/PhysRevE.73.036618} {\bibfield
  {journal} {\bibinfo  {journal} {Phys. Rev. E}\ }\textbf {\bibinfo {volume}
  {73}},\ \bibinfo {pages} {036618} (\bibinfo {year} {2006})}\BibitemShut
  {NoStop}%
\bibitem [{\citenamefont
  {Rangarajan}(1998)}]{rangarajan1998kolmogorov-arnold-moser}%
  \BibitemOpen
  \bibfield  {author} {\bibinfo {author} {\bibfnamefont {G.}~\bibnamefont
  {Rangarajan}},\ }\href@noop {} {\bibfield  {journal} {\bibinfo  {journal}
  {Resonance}\ }\textbf {\bibinfo {volume} {3}},\ \bibinfo {pages} {43}
  (\bibinfo {year} {1998})}\BibitemShut {NoStop}%
\bibitem [{\citenamefont {{Gallavotti}}(2008)}]{2008LNP728G}%
  \BibitemOpen
  \bibinfo {editor} {\bibfnamefont {G.}~\bibnamefont {{Gallavotti}}},\ ed.,\
  \href {\doibase 10.1007/978-3-540-72995-2} {\emph {\bibinfo {title} {The
  Fermi-Pasta-Ulam Problem Lecture Notes in Physics}}},\ \bibinfo {series}
  {Berlin Springer Verlag}, Vol.\ \bibinfo {volume} {728}\ (\bibinfo {year}
  {2008})\BibitemShut {NoStop}%
\bibitem [{\citenamefont {Ford}\ and\ \citenamefont
  {Lunsford}(1970)}]{PhysRevA.1.59}%
  \BibitemOpen
  \bibfield  {author} {\bibinfo {author} {\bibfnamefont {J.}~\bibnamefont
  {Ford}}\ and\ \bibinfo {author} {\bibfnamefont {G.~H.}\ \bibnamefont
  {Lunsford}},\ }\href {\doibase 10.1103/PhysRevA.1.59} {\bibfield  {journal}
  {\bibinfo  {journal} {Phys. Rev. A}\ }\textbf {\bibinfo {volume} {1}},\
  \bibinfo {pages} {59} (\bibinfo {year} {1970})}\BibitemShut {NoStop}%
\bibitem [{\citenamefont {Tuck}\ and\ \citenamefont
  {Menzel}(1972)}]{TUCK1972399}%
  \BibitemOpen
  \bibfield  {author} {\bibinfo {author} {\bibfnamefont {J.}~\bibnamefont
  {Tuck}}\ and\ \bibinfo {author} {\bibfnamefont {M.}~\bibnamefont {Menzel}},\
  }\href {\doibase 10.1016/0001-8708(72)90024-2} {\bibfield  {journal}
  {\bibinfo  {journal} {Adv. Math.}\ }\textbf {\bibinfo {volume} {9}},\
  \bibinfo {pages} {399 } (\bibinfo {year} {1972})}\BibitemShut {NoStop}%
\bibitem [{\citenamefont {Bivins}\ \emph {et~al.}(1973)\citenamefont {Bivins},
  \citenamefont {Metropolis},\ and\ \citenamefont {Pasta}}]{BIVINS197365}%
  \BibitemOpen
  \bibfield  {author} {\bibinfo {author} {\bibfnamefont {R.}~\bibnamefont
  {Bivins}}, \bibinfo {author} {\bibfnamefont {N.}~\bibnamefont {Metropolis}},
  \ and\ \bibinfo {author} {\bibfnamefont {J.~R.}\ \bibnamefont {Pasta}},\
  }\href {\doibase 10.1016/0021-9991(73)90169-1} {\bibfield  {journal}
  {\bibinfo  {journal} {J. Comput. Phys.}\ }\textbf {\bibinfo {volume} {12}},\
  \bibinfo {pages} {65 } (\bibinfo {year} {1973})}\BibitemShut {NoStop}%
\bibitem [{\citenamefont {Sholl}(1990)}]{SHOLL1990253}%
  \BibitemOpen
  \bibfield  {author} {\bibinfo {author} {\bibfnamefont {D.}~\bibnamefont
  {Sholl}},\ }\href {\doibase 10.1016/0375-9601(90)90424-M} {\bibfield
  {journal} {\bibinfo  {journal} {Phys. Lett. A}\ }\textbf {\bibinfo {volume}
  {149}},\ \bibinfo {pages} {253 } (\bibinfo {year} {1990})}\BibitemShut
  {NoStop}%
\bibitem [{\citenamefont {Ford}(1992)}]{FORD1992271}%
  \BibitemOpen
  \bibfield  {author} {\bibinfo {author} {\bibfnamefont {J.}~\bibnamefont
  {Ford}},\ }\href {\doibase 10.1016/0370-1573(92)90116-H} {\bibfield
  {journal} {\bibinfo  {journal} {Phys. Rep.}\ }\textbf {\bibinfo {volume}
  {213}},\ \bibinfo {pages} {271 } (\bibinfo {year} {1992})}\BibitemShut
  {NoStop}%
\bibitem [{\citenamefont {Zaslavsky}(2005)}]{doi:10.1063/1.1858115}%
  \BibitemOpen
  \bibfield  {author} {\bibinfo {author} {\bibfnamefont {G.~M.}\ \bibnamefont
  {Zaslavsky}},\ }\href {\doibase 10.1063/1.1858115} {\bibfield  {journal}
  {\bibinfo  {journal} {Chaos}\ }\textbf {\bibinfo {volume} {15}},\ \bibinfo
  {pages} {015103} (\bibinfo {year} {2005})}\BibitemShut {NoStop}%
\bibitem [{\citenamefont {Carati}\ \emph {et~al.}(2005)\citenamefont {Carati},
  \citenamefont {Galgani},\ and\ \citenamefont
  {Giorgilli}}]{doi:10.1063/1.1861264}%
  \BibitemOpen
  \bibfield  {author} {\bibinfo {author} {\bibfnamefont {A.}~\bibnamefont
  {Carati}}, \bibinfo {author} {\bibfnamefont {L.}~\bibnamefont {Galgani}}, \
  and\ \bibinfo {author} {\bibfnamefont {A.}~\bibnamefont {Giorgilli}},\ }\href
  {\doibase 10.1063/1.1861264} {\bibfield  {journal} {\bibinfo  {journal}
  {Chaos}\ }\textbf {\bibinfo {volume} {15}},\ \bibinfo {pages} {015105}
  (\bibinfo {year} {2005})}\BibitemShut {NoStop}%
\bibitem [{\citenamefont {Danieli}\ \emph {et~al.}(2017)\citenamefont
  {Danieli}, \citenamefont {Campbell},\ and\ \citenamefont
  {Flach}}]{PhysRevE.95.060202}%
  \BibitemOpen
  \bibfield  {author} {\bibinfo {author} {\bibfnamefont {C.}~\bibnamefont
  {Danieli}}, \bibinfo {author} {\bibfnamefont {D.~K.}\ \bibnamefont
  {Campbell}}, \ and\ \bibinfo {author} {\bibfnamefont {S.}~\bibnamefont
  {Flach}},\ }\href {\doibase 10.1103/PhysRevE.95.060202} {\bibfield  {journal}
  {\bibinfo  {journal} {Phys. Rev. E}\ }\textbf {\bibinfo {volume} {95}},\
  \bibinfo {pages} {060202} (\bibinfo {year} {2017})}\BibitemShut {NoStop}%
\bibitem [{\citenamefont {Berchialla}\ \emph {et~al.}(2004)\citenamefont
  {Berchialla}, \citenamefont {Giorgilli},\ and\ \citenamefont
  {Paleari}}]{BERCHIALLA2004167}%
  \BibitemOpen
  \bibfield  {author} {\bibinfo {author} {\bibfnamefont {L.}~\bibnamefont
  {Berchialla}}, \bibinfo {author} {\bibfnamefont {A.}~\bibnamefont
  {Giorgilli}}, \ and\ \bibinfo {author} {\bibfnamefont {S.}~\bibnamefont
  {Paleari}},\ }\href {\doibase 10.1016/j.physleta.2003.11.052} {\bibfield
  {journal} {\bibinfo  {journal} {Phys. Lett. A}\ }\textbf {\bibinfo {volume}
  {321}},\ \bibinfo {pages} {167 } (\bibinfo {year} {2004})}\BibitemShut
  {NoStop}%
\bibitem [{\citenamefont {DeLuca}\ \emph {et~al.}(1995)\citenamefont {DeLuca},
  \citenamefont {Lichtenberg},\ and\ \citenamefont {Ruffo}}]{PhysRevE.51.2877}%
  \BibitemOpen
  \bibfield  {author} {\bibinfo {author} {\bibfnamefont {J.}~\bibnamefont
  {DeLuca}}, \bibinfo {author} {\bibfnamefont {A.~J.}\ \bibnamefont
  {Lichtenberg}}, \ and\ \bibinfo {author} {\bibfnamefont {S.}~\bibnamefont
  {Ruffo}},\ }\href {\doibase 10.1103/PhysRevE.51.2877} {\bibfield  {journal}
  {\bibinfo  {journal} {Phys. Rev. E}\ }\textbf {\bibinfo {volume} {51}},\
  \bibinfo {pages} {2877} (\bibinfo {year} {1995})}\BibitemShut {NoStop}%
\bibitem [{\citenamefont {DeLuca}\ \emph {et~al.}(1999)\citenamefont {DeLuca},
  \citenamefont {Lichtenberg},\ and\ \citenamefont {Ruffo}}]{PhysRevE.60.3781}%
  \BibitemOpen
  \bibfield  {author} {\bibinfo {author} {\bibfnamefont {J.}~\bibnamefont
  {DeLuca}}, \bibinfo {author} {\bibfnamefont {A.~J.}\ \bibnamefont
  {Lichtenberg}}, \ and\ \bibinfo {author} {\bibfnamefont {S.}~\bibnamefont
  {Ruffo}},\ }\href {\doibase 10.1103/PhysRevE.60.3781} {\bibfield  {journal}
  {\bibinfo  {journal} {Phys. Rev. E}\ }\textbf {\bibinfo {volume} {60}},\
  \bibinfo {pages} {3781} (\bibinfo {year} {1999})}\BibitemShut {NoStop}%
\bibitem [{\citenamefont {Benettin}\ and\ \citenamefont
  {Ponno}(2011)}]{Benettin2011}%
  \BibitemOpen
  \bibfield  {author} {\bibinfo {author} {\bibfnamefont {G.}~\bibnamefont
  {Benettin}}\ and\ \bibinfo {author} {\bibfnamefont {A.}~\bibnamefont
  {Ponno}},\ }\href {\doibase 10.1007/s10955-011-0277-9} {\bibfield  {journal}
  {\bibinfo  {journal} {J. Stat. Phys.}\ }\textbf {\bibinfo {volume} {144}},\
  \bibinfo {pages} {793} (\bibinfo {year} {2011})}\BibitemShut {NoStop}%
\bibitem [{\citenamefont {Benettin}\ \emph {et~al.}(2013)\citenamefont
  {Benettin}, \citenamefont {Christodoulidi},\ and\ \citenamefont
  {Ponno}}]{Benettin2013}%
  \BibitemOpen
  \bibfield  {author} {\bibinfo {author} {\bibfnamefont {G.}~\bibnamefont
  {Benettin}}, \bibinfo {author} {\bibfnamefont {H.}~\bibnamefont
  {Christodoulidi}}, \ and\ \bibinfo {author} {\bibfnamefont {A.}~\bibnamefont
  {Ponno}},\ }\href {\doibase 10.1007/s10955-013-0760-6} {\bibfield  {journal}
  {\bibinfo  {journal} {J. Stat. Phys.}\ }\textbf {\bibinfo {volume} {152}},\
  \bibinfo {pages} {195} (\bibinfo {year} {2013})}\BibitemShut {NoStop}%
\bibitem [{\citenamefont {Onorato}\ \emph {et~al.}(2015)\citenamefont
  {Onorato}, \citenamefont {Vozella}, \citenamefont {Proment},\ and\
  \citenamefont {Lvov}}]{Onorato4208}%
  \BibitemOpen
  \bibfield  {author} {\bibinfo {author} {\bibfnamefont {M.}~\bibnamefont
  {Onorato}}, \bibinfo {author} {\bibfnamefont {L.}~\bibnamefont {Vozella}},
  \bibinfo {author} {\bibfnamefont {D.}~\bibnamefont {Proment}}, \ and\
  \bibinfo {author} {\bibfnamefont {Y.~V.}\ \bibnamefont {Lvov}},\ }\href
  {\doibase 10.1073/pnas.1404397112} {\bibfield  {journal} {\bibinfo  {journal}
  {Proc. Natl. Acad. Sci. U.S.A.}\ }\textbf {\bibinfo {volume} {112}},\
  \bibinfo {pages} {4208} (\bibinfo {year} {2015})}\BibitemShut {NoStop}%
\bibitem [{\citenamefont {Lvov}\ and\ \citenamefont
  {Onorato}(2018)}]{PhysRevLett.120.144301}%
  \BibitemOpen
  \bibfield  {author} {\bibinfo {author} {\bibfnamefont {Y.~V.}\ \bibnamefont
  {Lvov}}\ and\ \bibinfo {author} {\bibfnamefont {M.}~\bibnamefont {Onorato}},\
  }\href {\doibase 10.1103/PhysRevLett.120.144301} {\bibfield  {journal}
  {\bibinfo  {journal} {Phys. Rev. Lett.}\ }\textbf {\bibinfo {volume} {120}},\
  \bibinfo {pages} {144301} (\bibinfo {year} {2018})}\BibitemShut {NoStop}%
\bibitem [{\citenamefont {Pistone}\ \emph {et~al.}(2018)\citenamefont
  {Pistone}, \citenamefont {Onorato},\ and\ \citenamefont
  {Chibbaro}}]{0295-5075-121-4-44003}%
  \BibitemOpen
  \bibfield  {author} {\bibinfo {author} {\bibfnamefont {L.}~\bibnamefont
  {Pistone}}, \bibinfo {author} {\bibfnamefont {M.}~\bibnamefont {Onorato}}, \
  and\ \bibinfo {author} {\bibfnamefont {S.}~\bibnamefont {Chibbaro}},\ }\href
  {http://stacks.iop.org/0295-5075/121/i=4/a=44003} {\bibfield  {journal}
  {\bibinfo  {journal} {EPL (Europhysics Letters)}\ }\textbf {\bibinfo {volume}
  {121}},\ \bibinfo {pages} {44003} (\bibinfo {year} {2018})}\BibitemShut
  {NoStop}%
\bibitem [{\citenamefont {Bustamante}\ \emph {et~al.}(2018)\citenamefont
  {Bustamante}, \citenamefont {Hutchinson}, \citenamefont {Lvov},\ and\
  \citenamefont {Onorato}}]{arXiv:1810.06902}%
  \BibitemOpen
  \bibfield  {author} {\bibinfo {author} {\bibfnamefont {M.~D.}\ \bibnamefont
  {Bustamante}}, \bibinfo {author} {\bibfnamefont {K.}~\bibnamefont
  {Hutchinson}}, \bibinfo {author} {\bibfnamefont {Y.~V.}\ \bibnamefont
  {Lvov}}, \ and\ \bibinfo {author} {\bibfnamefont {M.}~\bibnamefont
  {Onorato}},\ }\href {https://arxiv.org/abs/1810.06902} {\bibfield  {journal}
  {\bibinfo  {journal} {arXiv:1810.06902}\ } (\bibinfo {year}
  {2018})}\BibitemShut {NoStop}%
\bibitem [{\citenamefont {{Zakharov}}\ \emph {et~al.}(1992)\citenamefont
  {{Zakharov}}, \citenamefont {{L'Vov}},\ and\ \citenamefont
  {{Falkovich}}}]{1992kstbookZ}%
  \BibitemOpen
  \bibfield  {author} {\bibinfo {author} {\bibfnamefont {V.~E.}\ \bibnamefont
  {{Zakharov}}}, \bibinfo {author} {\bibfnamefont {V.~S.}\ \bibnamefont
  {{L'Vov}}}, \ and\ \bibinfo {author} {\bibfnamefont {G.}~\bibnamefont
  {{Falkovich}}},\ }\href {\doibase 10.1007/978-3-642-50052-7} {\emph {\bibinfo
  {title} {Kolmogorov spectra of turbulence I.~Wave turbulence.}}},\ Springer,
  Berlin (Germany),\ (\bibinfo {year} {1992})\BibitemShut {NoStop}%
\bibitem [{\citenamefont {Majda}\ \emph {et~al.}(1997)\citenamefont {Majda},
  \citenamefont {McLaughlin},\ and\ \citenamefont {Tabak}}]{Majda1997}%
  \BibitemOpen
  \bibfield  {author} {\bibinfo {author} {\bibfnamefont {A.~J.}\ \bibnamefont
  {Majda}}, \bibinfo {author} {\bibfnamefont {D.~W.}\ \bibnamefont
  {McLaughlin}}, \ and\ \bibinfo {author} {\bibfnamefont {E.~G.}\ \bibnamefont
  {Tabak}},\ }\href {\doibase 10.1007/BF02679124} {\bibfield  {journal}
  {\bibinfo  {journal} {J. Nonlinear Sci.}\ }\textbf {\bibinfo {volume} {7}},\
  \bibinfo {pages} {9} (\bibinfo {year} {1997})}\BibitemShut {NoStop}%
\bibitem [{\citenamefont {Zakharov}\ \emph {et~al.}(2001)\citenamefont
  {Zakharov}, \citenamefont {Guyenne}, \citenamefont {Pushkarev},\ and\
  \citenamefont {Dias}}]{ZAKHAROV2001573}%
  \BibitemOpen
  \bibfield  {author} {\bibinfo {author} {\bibfnamefont {V.}~\bibnamefont
  {Zakharov}}, \bibinfo {author} {\bibfnamefont {P.}~\bibnamefont {Guyenne}},
  \bibinfo {author} {\bibfnamefont {A.}~\bibnamefont {Pushkarev}}, \ and\
  \bibinfo {author} {\bibfnamefont {F.}~\bibnamefont {Dias}},\ }\href {\doibase
  10.1016/S0167-2789(01)00194-4} {\bibfield  {journal} {\bibinfo  {journal}
  {Physica D}\ }\textbf {\bibinfo {volume} {152-153}},\ \bibinfo {pages} {573 }
  (\bibinfo {year} {2001})}\BibitemShut {NoStop}%
\bibitem [{\citenamefont {Zakharov}\ \emph {et~al.}(2004)\citenamefont
  {Zakharov}, \citenamefont {Dias},\ and\ \citenamefont
  {Pushkarev}}]{ZAKHAROV20041}%
  \BibitemOpen
  \bibfield  {author} {\bibinfo {author} {\bibfnamefont {V.}~\bibnamefont
  {Zakharov}}, \bibinfo {author} {\bibfnamefont {F.}~\bibnamefont {Dias}}, \
  and\ \bibinfo {author} {\bibfnamefont {A.}~\bibnamefont {Pushkarev}},\ }\href
  {\doibase 10.1016/j.physrep.2004.04.002} {\bibfield  {journal} {\bibinfo
  {journal} {Phys. Rep.}\ }\textbf {\bibinfo {volume} {398}},\ \bibinfo {pages}
  {1 } (\bibinfo {year} {2004})}\BibitemShut {NoStop}%
\bibitem [{\citenamefont {{Nazarenko}}(2011)}]{2011LNP825N}%
  \BibitemOpen
  \bibinfo {editor} {\bibfnamefont {S.}~\bibnamefont {{Nazarenko}}},\ ed.,\
  \href {\doibase 10.1007/978-3-642-15942-8} {\emph {\bibinfo {title} {Wave
  Turbulence, Lecture Notes in Physics}}},\ \bibinfo {series} {Berlin Springer
  Verlag}, Vol.\ \bibinfo {volume} {825}\ (\bibinfo {year} {2011})\BibitemShut
  {NoStop}%
\bibitem [{\citenamefont {{Sagaut}}\ and\ \citenamefont
  {{Cambon}}(2018)}]{2018htdbookZ}%
  \BibitemOpen
  \bibfield  {author} {\bibinfo {author} {\bibfnamefont {P.}~\bibnamefont
  {{Sagaut}}}\ and\ \bibinfo {author} {\bibfnamefont {C.}~\bibnamefont
  {{Cambon}}},\ }\href {\doibase 10.1007/978-3-319-73162-9} {\emph {\bibinfo
  {title} {Homogeneous Turbulence Dynamics.}}},\ Springer International
  Publishing AG,\ (\bibinfo {year} {2018})\BibitemShut {NoStop}%
\bibitem [{\citenamefont {Fu}\ \emph {et~al.}(2018)\citenamefont {Fu},
  \citenamefont {Zhang},\ and\ \citenamefont {Zhao}}]{Our2018}%
  \BibitemOpen
  \bibfield  {author} {\bibinfo {author} {\bibfnamefont {W.}~\bibnamefont
  {Fu}}, \bibinfo {author} {\bibfnamefont {Y.}~\bibnamefont {Zhang}}, \ and\
  \bibinfo {author} {\bibfnamefont {H.}~\bibnamefont {Zhao}},\ }\href
  {https://arxiv.org/abs/1811.05697} {\bibfield  {journal} {\bibinfo  {journal}
  {arXiv:1811.05697}\ } (\bibinfo {year} {2018})}\BibitemShut {NoStop}%
\bibitem [{\citenamefont {Toda}(1967)}]{1967431}%
  \BibitemOpen
  \bibfield  {author} {\bibinfo {author} {\bibfnamefont {M.}~\bibnamefont
  {Toda}},\ }\href {\doibase 10.1143/JPSJ.22.431} {\bibfield  {journal}
  {\bibinfo  {journal} {J. Phys. Soc. Jpn.}\ }\textbf {\bibinfo {volume}
  {22}},\ \bibinfo {pages} {431} (\bibinfo {year} {1967})}\BibitemShut
  {NoStop}%
\bibitem [{\citenamefont {Ferguson}\ \emph {et~al.}(1982)\citenamefont
  {Ferguson}, \citenamefont {Flaschka},\ and\ \citenamefont
  {McLaughlin}}]{FERGUSON1982157}%
  \BibitemOpen
  \bibfield  {author} {\bibinfo {author} {\bibfnamefont {W.}~\bibnamefont
  {Ferguson}}, \bibinfo {author} {\bibfnamefont {H.}~\bibnamefont {Flaschka}},
  \ and\ \bibinfo {author} {\bibfnamefont {D.}~\bibnamefont {McLaughlin}},\
  }\href {\doibase 10.1016/0021-9991(82)90116-4} {\bibfield  {journal}
  {\bibinfo  {journal} {J. Computat. Phys.}\ }\textbf {\bibinfo {volume}
  {45}},\ \bibinfo {pages} {157 } (\bibinfo {year} {1982})}\BibitemShut
  {NoStop}%
\bibitem [{\citenamefont {Casetti}\ \emph {et~al.}(1997)\citenamefont
  {Casetti}, \citenamefont {Cerruti-Sola}, \citenamefont {Pettini},\ and\
  \citenamefont {Cohen}}]{PhysRevE.55.6566}%
  \BibitemOpen
  \bibfield  {author} {\bibinfo {author} {\bibfnamefont {L.}~\bibnamefont
  {Casetti}}, \bibinfo {author} {\bibfnamefont {M.}~\bibnamefont
  {Cerruti-Sola}}, \bibinfo {author} {\bibfnamefont {M.}~\bibnamefont
  {Pettini}}, \ and\ \bibinfo {author} {\bibfnamefont {E.~G.~D.}\ \bibnamefont
  {Cohen}},\ }\href {\doibase 10.1103/PhysRevE.55.6566} {\bibfield  {journal}
  {\bibinfo  {journal} {Phys. Rev. E}\ }\textbf {\bibinfo {volume} {55}},\
  \bibinfo {pages} {6566} (\bibinfo {year} {1997})}\BibitemShut {NoStop}%
\bibitem [{\citenamefont {Cerruti-Sola}\ \emph {et~al.}(2000)\citenamefont
  {Cerruti-Sola}, \citenamefont {Pettini},\ and\ \citenamefont
  {Cohen}}]{PhysRevE.62.6078}%
  \BibitemOpen
  \bibfield  {author} {\bibinfo {author} {\bibfnamefont {M.}~\bibnamefont
  {Cerruti-Sola}}, \bibinfo {author} {\bibfnamefont {M.}~\bibnamefont
  {Pettini}}, \ and\ \bibinfo {author} {\bibfnamefont {E.~G.~D.}\ \bibnamefont
  {Cohen}},\ }\href {\doibase 10.1103/PhysRevE.62.6078} {\bibfield  {journal}
  {\bibinfo  {journal} {Phys. Rev. E}\ }\textbf {\bibinfo {volume} {62}},\
  \bibinfo {pages} {6078} (\bibinfo {year} {2000})}\BibitemShut {NoStop}%
\bibitem [{\citenamefont {Benettin}\ \emph {et~al.}(2018)\citenamefont
  {Benettin}, \citenamefont {Pasquali},\ and\ \citenamefont
  {Ponno}}]{Benettin2018}%
  \BibitemOpen
  \bibfield  {author} {\bibinfo {author} {\bibfnamefont {G.}~\bibnamefont
  {Benettin}}, \bibinfo {author} {\bibfnamefont {S.}~\bibnamefont {Pasquali}},
  \ and\ \bibinfo {author} {\bibfnamefont {A.}~\bibnamefont {Ponno}},\ }\href
  {\doibase 10.1007/s10955-018-2017-x} {\bibfield  {journal} {\bibinfo
  {journal} {J. Stat. Phys.}\ }\textbf {\bibinfo {volume} {171}},\ \bibinfo
  {pages} {521} (\bibinfo {year} {2018})}\BibitemShut {NoStop}%
\bibitem [{\citenamefont {Zakharov}\ and\ \citenamefont
  {Schulman}(1988)}]{ZAKHAROV1988283}%
  \BibitemOpen
  \bibfield  {author} {\bibinfo {author} {\bibfnamefont {V.}~\bibnamefont
  {Zakharov}}\ and\ \bibinfo {author} {\bibfnamefont {E.}~\bibnamefont
  {Schulman}},\ }\href {\doibase 10.1016/0167-2789(88)90033-4} {\bibfield
  {journal} {\bibinfo  {journal} {Physica D}\ }\textbf {\bibinfo {volume}
  {29}},\ \bibinfo {pages} {283 } (\bibinfo {year} {1988})}\BibitemShut
  {NoStop}%
\bibitem [{\citenamefont {Livi}\ \emph {et~al.}(1985)\citenamefont {Livi},
  \citenamefont {Pettini}, \citenamefont {Ruffo}, \citenamefont
  {Sparpaglione},\ and\ \citenamefont {Vulpiani}}]{PhysRevA.31.1039}%
  \BibitemOpen
  \bibfield  {author} {\bibinfo {author} {\bibfnamefont {R.}~\bibnamefont
  {Livi}}, \bibinfo {author} {\bibfnamefont {M.}~\bibnamefont {Pettini}},
  \bibinfo {author} {\bibfnamefont {S.}~\bibnamefont {Ruffo}}, \bibinfo
  {author} {\bibfnamefont {M.}~\bibnamefont {Sparpaglione}}, \ and\ \bibinfo
  {author} {\bibfnamefont {A.}~\bibnamefont {Vulpiani}},\ }\href {\doibase
  10.1103/PhysRevA.31.1039} {\bibfield  {journal} {\bibinfo  {journal} {Phys.
  Rev. A}\ }\textbf {\bibinfo {volume} {31}},\ \bibinfo {pages} {1039}
  (\bibinfo {year} {1985})}\BibitemShut {NoStop}%
\bibitem [{\citenamefont {Goedde}\ \emph {et~al.}(1992)\citenamefont {Goedde},
  \citenamefont {Lichtenberg},\ and\ \citenamefont
  {Lieberman}}]{GOEDDE1992200}%
  \BibitemOpen
  \bibfield  {author} {\bibinfo {author} {\bibfnamefont {C.}~\bibnamefont
  {Goedde}}, \bibinfo {author} {\bibfnamefont {A.}~\bibnamefont {Lichtenberg}},
  \ and\ \bibinfo {author} {\bibfnamefont {M.}~\bibnamefont {Lieberman}},\
  }\href {\doibase 10.1016/0167-2789(92)90216-A} {\bibfield  {journal}
  {\bibinfo  {journal} {Physica D}\ }\textbf {\bibinfo {volume} {59}},\
  \bibinfo {pages} {200 } (\bibinfo {year} {1992})}\BibitemShut {NoStop}%
\bibitem [{\citenamefont {Yoshida}(1990)}]{YOSHIDA1990262}%
  \BibitemOpen
  \bibfield  {author} {\bibinfo {author} {\bibfnamefont {H.}~\bibnamefont
  {Yoshida}},\ }\href {\doibase 10.1016/0375-9601(90)90092-3} {\bibfield
  {journal} {\bibinfo  {journal} {Phys. Lett. A}\ }\textbf {\bibinfo {volume}
  {150}},\ \bibinfo {pages} {262 } (\bibinfo {year} {1990})}\BibitemShut
  {NoStop}%
\bibitem [{\citenamefont {Benettin}\ \emph {et~al.}(2009)\citenamefont
  {Benettin}, \citenamefont {Livi},\ and\ \citenamefont
  {Ponno}}]{Benettin2009}%
  \BibitemOpen
  \bibfield  {author} {\bibinfo {author} {\bibfnamefont {G.}~\bibnamefont
  {Benettin}}, \bibinfo {author} {\bibfnamefont {R.}~\bibnamefont {Livi}}, \
  and\ \bibinfo {author} {\bibfnamefont {A.}~\bibnamefont {Ponno}},\ }\href
  {\doibase 10.1007/s10955-008-9660-6} {\bibfield  {journal} {\bibinfo
  {journal} {J. Stat. Phys.}\ }\textbf {\bibinfo {volume} {135}},\ \bibinfo
  {pages} {873} (\bibinfo {year} {2009})}\BibitemShut {NoStop}%
\bibitem [{\citenamefont {Ponno}\ \emph {et~al.}(2011)\citenamefont {Ponno},
  \citenamefont {Christodoulidi}, \citenamefont {Skokos},\ and\ \citenamefont
  {Flach}}]{doi:10.1063/1.3658620}%
  \BibitemOpen
  \bibfield  {author} {\bibinfo {author} {\bibfnamefont {A.}~\bibnamefont
  {Ponno}}, \bibinfo {author} {\bibfnamefont {H.}~\bibnamefont
  {Christodoulidi}}, \bibinfo {author} {\bibfnamefont {C.}~\bibnamefont
  {Skokos}}, \ and\ \bibinfo {author} {\bibfnamefont {S.}~\bibnamefont
  {Flach}},\ }\href {\doibase 10.1063/1.3658620} {\bibfield  {journal}
  {\bibinfo  {journal} {Chaos}\ }\textbf {\bibinfo {volume} {21}},\ \bibinfo
  {pages} {043127} (\bibinfo {year} {2011})}\BibitemShut {NoStop}%
\bibitem [{\citenamefont {Goldfriend}\ and\ \citenamefont
  {Kurchan}(2018)}]{arXiv:1810.06121}%
  \BibitemOpen
  \bibfield  {author} {\bibinfo {author} {\bibfnamefont {T.}~\bibnamefont
  {Goldfriend}}\ and\ \bibinfo {author} {\bibfnamefont {J.}~\bibnamefont
  {Kurchan}},\ }\href {https://arxiv.org/abs/1810.06121v1} {\bibfield
  {journal} {\bibinfo  {journal} {arXiv:1810.06121}\ } (\bibinfo {year}
  {2018})}\BibitemShut {NoStop}%
\bibitem [{\citenamefont {Zhang}\ \emph {et~al.}(2016)\citenamefont {Zhang},
  \citenamefont {Tang},\ and\ \citenamefont {Tong}}]{PhysRevE.93.022216}%
  \BibitemOpen
  \bibfield  {author} {\bibinfo {author} {\bibfnamefont {Z.}~\bibnamefont
  {Zhang}}, \bibinfo {author} {\bibfnamefont {C.}~\bibnamefont {Tang}}, \ and\
  \bibinfo {author} {\bibfnamefont {P.}~\bibnamefont {Tong}},\ }\href {\doibase
  10.1103/PhysRevE.93.022216} {\bibfield  {journal} {\bibinfo  {journal} {Phys.
  Rev. E}\ }\textbf {\bibinfo {volume} {93}},\ \bibinfo {pages} {022216}
  (\bibinfo {year} {2016})}\BibitemShut {NoStop}%
\end{thebibliography}%

\end{document}